\documentclass{iaus}
\usepackage{graphicx}

\title[Massive galaxies and Dark matter Haloes] 
{Breaking down the link between luminous and dark matter in massive galaxies}

\author[S. Foucaud \& C. J. Conselice]   
{S\'ebastien Foucaud$^1$ \& Christopher J. Conselice$^2$}

\affiliation{$^1$National Taiwan Normal University, Taiwan \\ email: {\tt foucaud@ntnu.edu.tw} \\
$^2$University of Nottingham, UK \\ email: {\tt conselice@nottingham.ac.uk} \\[\affilskip]}

\pubyear{2010}
\volume{277}  
\jname{Tracing the Origin of Galaxies in the Land of our Ancestors}
\editors{Carignan, C., Combes, F. \& Freeman, K. C., eds.}
\begin{document}

\maketitle
\begin{abstract}

We present a study on the clustering of a stellar mass selected sample of galaxies with stellar masses $M_* > 10^{10} M_{\odot}$ at redshifts $0.4 < z < 2.0$, taken from the Palomar Observatory Wide-field Infrared Survey. We examine the clustering properties of these stellar mass selected samples as a function of redshift and stellar mass, and find that galaxies with high stellar masses have a progressively higher clustering strength than galaxies with lower stellar masses. We also find that galaxies within a fixed stellar mass range have a higher clustering strength at higher redshifts. We further estimate the average total masses of the dark matter haloes hosting these stellar-mass selected galaxies. For all galaxies in our sample the stellar-mass-to-total-mass ratio is always lower than the universal baryonic mass fraction and the stellar-mass-to-total-mass ratio is strongly correlated with the halo masses for central galaxies, such that more massive haloes contain a lower fraction of their mass in the form of stars. The remaining baryonic mass is included partially in stars within satellite galaxies in these haloes, and as diffuse hot and warm gas. We also find that, at a fixed stellar mass, the stellar-to-total-mass ratio increases at lower redshifts. This suggests that galaxies at a fixed stellar mass form later in lower mass dark matter haloes, and earlier in massive haloes. We interpret this as a `halo downsizing' effect.

\end{abstract}

\firstsection 
\section{Introduction}

Stellar masses are now becoming a standard measure for galaxies, and
are being used to trace the evolution of the galaxy population in
terms of star formation rates and morphologies (e.g. Bundy et al. 2005; Conselice et al. 2008; Cowie \& Barger 2008).
However, stellar mass only traces one aspect of galaxy mass, 
and ideally and ultimately, we aim to measure
galaxy total masses, that include contributions from stars, gas, and
dark matter.  Galaxies are believed to be hosted by massive dark
matter haloes that make up more than 85\% of their total mass,
and thus clearly tracing the co-evolution of galaxies and their haloes
is a major and important goal. One very powerful method for measuring the total masses of galaxies is
to measure their clustering.  Clustering measurements are independent
of photometric properties, and as such they can be used to highlight
fundamental properties of galaxy populations without assumptions
concerning stellar populations or mass profiles. Halo
clustering is a strong function of halo mass, with more massive haloes
more strongly clustered, providing a method to study the relationship
between galaxy properties and dark matter halo masses.

We present the first general study of the clustering
properties for a stellar mass selected sample of galaxies up to
$z\sim2$.  We carry this out by measuring the correlation length and amplitude
for galaxies selected with stellar masses $M_*>10^{10}M_{\odot}\,$ within
the Palomar Observatory Wide-field Infrared Survey (POWIR).

\section{The Palomar/DEEP2 Survey}

The Palomar Observatory Wide-Field survey (Bundy et al. 2006; Conselice et al. 2007) was designed to obtain deep Near-Infrared photometry over a 1.5deg$^2$ area, using the Wide Field Infrared Camera (WIRC) on the Palomar 5m telescope. It covers the Extended Groth Strip (Davis et al. 2007) and three other fields covered by the DEEP2 survey (Davis et al. 2003). Optical imaging from the Canada-France-Hawaii Telescope cover all fields, using the CFH12k camera in B-, R- and I-band (Coil et al. 2004). DEIMOS spectroscopy was acquired as part of the DEEP2 survey on the Keck Telescope. Around 20\% of our K-band selected galaxies have a secure redshift (up to z=1.4). For the galaxies without spectroscopic redshift, we have determined their photometric redshifts based on our BRIJK photometry and our spectroscopically confirmed sample.

Using our photometry we also derived stellar mass measures for our sample of galaxies (Bundy et al. 2006; Conselice et al. 2007). Based on models from Bruzual \& Charlot (2003), we derived accurate stellar masses in the range $10^{10.0} M_{\odot}<M_*<10^{12.0} M_{\odot}$ over our range of redshifts with an error of 0.2-0.3 dex.  In this work, we study the average halo masses of samples of galaxies selected by their stellar-masses and redshifts.

\section{Clustering properties of mass selected samples}

In order to link these galaxies with their environment and their large scale structure, we quantify their distribution in the sky at different scales. We first measure the 2-point angular correlation function $\omega(\theta)$ for our sample using the Landy \& Szalay (1993) estimator, as shown in Figure~1(a). In order to fit the clustering reliably, we avoid the small-scale excesses due to possible multiple galaxy occupation of a single dark matter halo. Assuming a Top-Hat redshift distribution for galaxies in each of our narrow redshift bins, we also derive the correlation length $r_0$ from the amplitude of the angular correlation using the relativistic Limber equation (Magliocchetti \& Maddox 1999).

As shown if Figure~1(b), we find that correlation lengths vary from $5 h^{-1}$Mpc  to $15 h^{-1}$Mpc  for galaxies selected by stellar masses with
$M_{*}>10^{10} M_{\odot}$. The correlation measurements of our different mass-selected samples indicate that the most massive galaxies are more clustered than less massive ones at all redshifts. Furthermore higher redshift samples are more clustered than lower redshift ones at a given mass range (Foucaud et al. 2010).

The standard Cold Dark Matter model predicts that at any redshift more massive dark matter haloes are on average more clustered than lower mass systems. To better understand where galaxies reside, their spatial distribution can be directly compared with the distribution of dark matter haloes as predicted by Mo \& White (2002).  The very strong clustering shown by $z>1$ massive galaxies implies that they are hosted by very massive dark matter halos, consistent with progenitors of present-day elliptical galaxies (Foucaud et al. 2010). Furthermore we can directly extract dark matter halo masses from the models, and compare directly this mass with the stellar mass of the hosted galaxy, deriving the stellar-to-dark matter mass fraction for our different samples. 

\begin{figure}[h]
\begin{center}
\includegraphics[width=3in, angle=90]{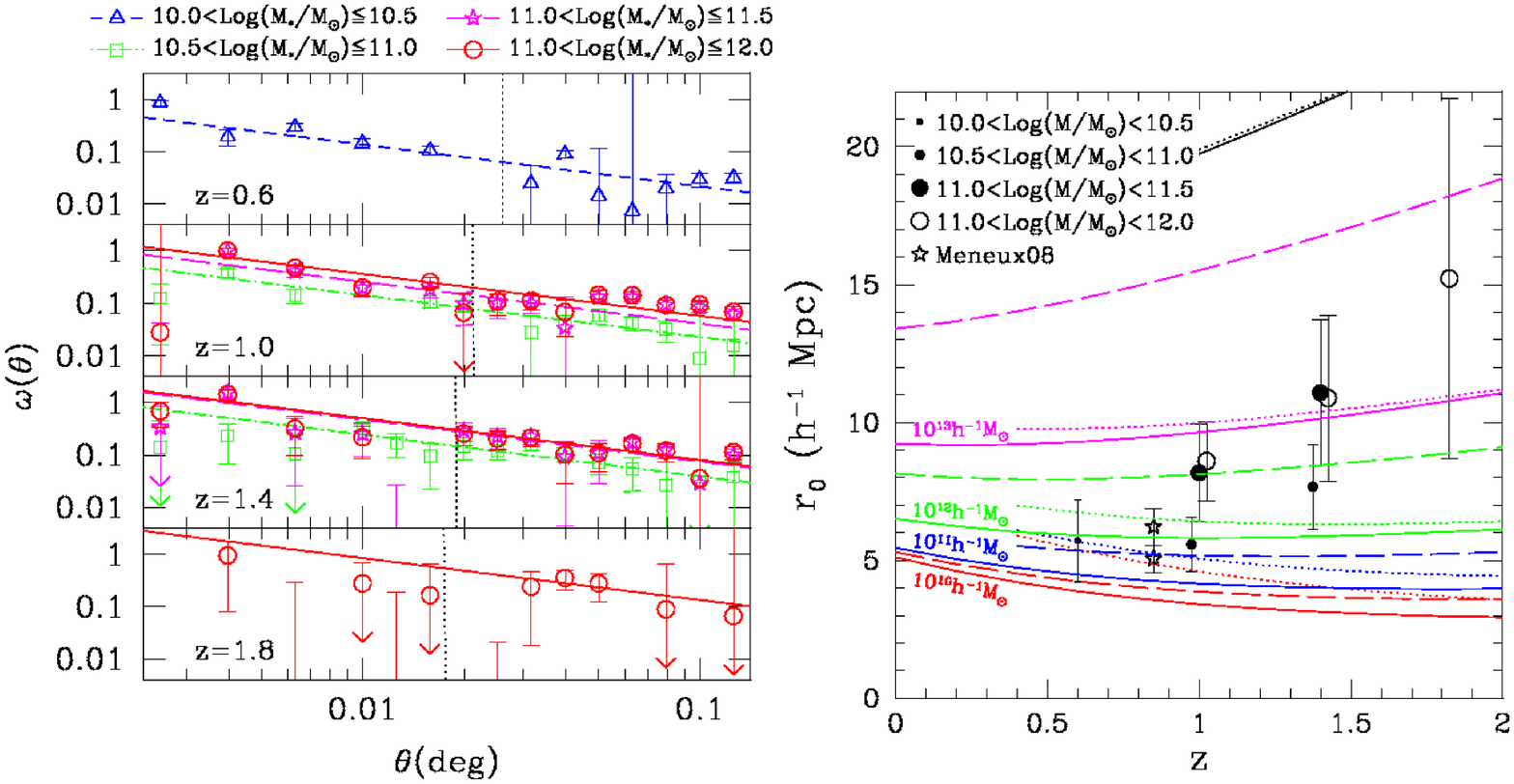} 
\vspace*{-1.0 cm}
\caption{{\bf (a)} Two-point correlation function of our mass-selected samples in different redshift ranges, in one of our field (Foucaud et al. 2010). The vertical dotted line corresponds to the lower limit at
    $\sim1 h^{-1}$Mpc of the range over which our data are fitted, in order to avoid any excess of pairs due to multiple halo occupation. {\bf (b)} Evolution of the spatial correlation length $r_0$ with redshift, for our different mass-selected samples (Foucaud et al. 2010). The measurements are compared with prediction of the evolution for dark matter haloes (Mo \& White 2002) and from the literature (Meneux et al. 2008). }
   \label{fig1}
\end{center}
\end{figure}

\section{Stellar-to-Dark Matter mass fraction}

In Figure 2(a), we compared our stellar-to-dark matter mass fractions with measurements from different studies in the literature based on various
methods to estimate the masses of dark matter haloes (rotation curves, galaxy-galaxy lensing, groups). As expected, the stellar-mass-to-total-mass ratio is far lower than the universal baryonic mass fraction, measured from WMAP5 (Komatsu et al. 2009). Furthermore we confirm at high redshift the tight correlation found between the stellar-to-dark matter mass fraction and dark matter halo masses in the higher mass regime. Overall, we find that, in all our redshift ranges, more massive systems have a lower ratio of
stellar-to-halo-mass. This implies that at $z>2$ galaxies in the centres of the most massive dark matter halo stop increasing their stellar mass while their halos keep growing.  This also implies a limit to how much stellar mass a galaxy can have, with a cut off
at a few times $10^{11} M_{\odot}$.   Similarly, a decreasing stellar mass fraction with increasing halo mass has been
observed in groups and clusters (e.g. Gonzalez et al. 2007), with the majority of the baryonic mass taking the form of galaxy satellites, while the remaining being under the form of gas.

As shown in Figure 2(b), the stellar-to-dark matter mass fraction also increases with lower redshift at a given stellar mass. This effect is likely linked with `contamination' of the massive sample over time by later forming galaxies hosted by less massive haloes. Since the stellar masses of the most massive galaxies do not increase much with time, due to 
a quenching of their star formation and a weak merging rate (Drory \& Alvarez 2008), the average mass of the haloes for these galaxies decreases, and
the stellar mass ratio increases. This is in agreement with a `halo downsizing' effect (Neistein et al. 2006), i.e. more massive haloes are formed earlier than lower mass haloes. 

\begin{figure}[h]
\begin{center}
\includegraphics[width=2.3in, angle=90]{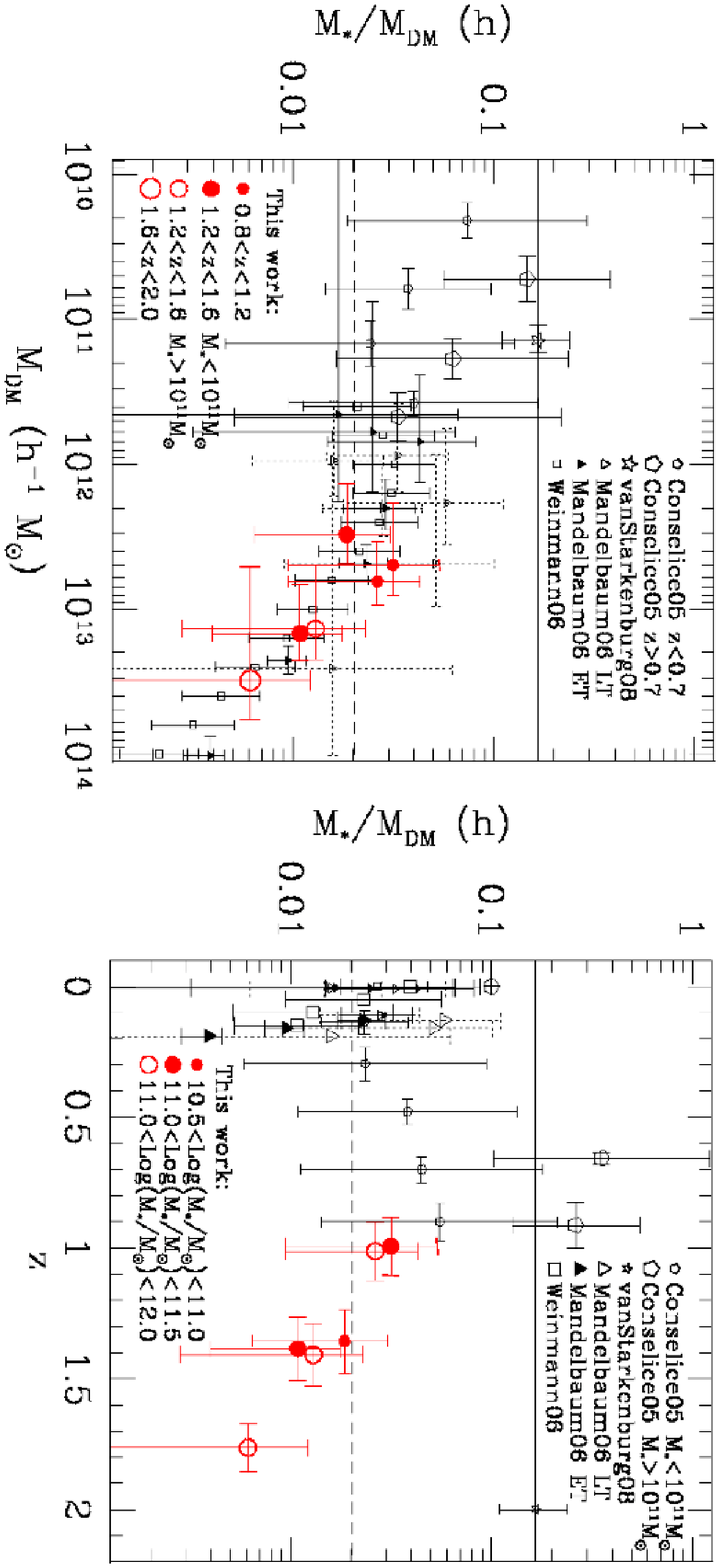} 
\caption{Evolution of the stellar to dark matter mass ratio with {\bf (a)} the mass of the dark matter halo in different redshift bins, and with {\bf (b)} the redshift in different stellar mass bins (Foucaud et al. 2010). We compared our measurements with literature (Conselice et al. 2005; Mandelbaum et al. 2006; van Starkenburg et al. 2008; Weinmann et al. 2006). The continuous line represents the baryonic fraction measured from  WMAP5 (Komatsu et al. 2009), and the dashed line the mean stellar fraction in the local Universe estimated by Cole et al. (2001).}
   \label{fig2}
\end{center}
\end{figure}

\section{Conclusions}

We have exploited the near infrared data set from the Palomar/DEEP2 survey to investigate the clustering evolution of the most massive galaxies observed from $z=2$ to $z=0$. By deriving the mass of the host dark matter haloes from clustering analyses, we show that the stellar fraction is lower in more massive dark matter haloes, and that more massive galaxies inhabit more massive dark matter haloes which form earlier, in agreement with downsizing effects. The global picture resulting from this study is that the most massive galaxies are formed early in massive dark matter halos. At some stage the evolution of these massive galaxies stops and henceforth evolves passively, while their halo keep growing in mass, with their baryonic mass in the form of satellites or gas.

\end{document}